\documentclass[%
 reprint,
 amsmath,amssymb,
 aps,
]{revtex4-1}

\usepackage[english]{babel}
\usepackage{amsmath}
\usepackage{graphicx}
\usepackage{dcolumn}
\usepackage{bm}
\usepackage{hyperref}
\usepackage{lipsum}

\begin{document}
\renewcommand{\figurename}{FIG.}

\preprint{APS/123-QED}

\title{Chiral optical Local Density of States in spiral plasmonic cavity}

\author{Aline Pham}\altaffiliation{Institut NEEL, CNRS and Universit\'e Grenoble Alpes, F-38000 Grenoble, France}
\author{Martin Berthel}\altaffiliation{Institut NEEL, CNRS and Universit\'e Grenoble Alpes, F-38000 Grenoble, France}
\author{Quanbo Jiang}\altaffiliation{Institut NEEL, CNRS and Universit\'e Grenoble Alpes, F-38000 Grenoble, France}
\author{Joel Bellessa}\altaffiliation{Institut Lumi\`ere Mati\`ere, UMR5306 Universit\'e Lyon 1-CNRS, Universit\'e de Lyon, 69622 Villeurbanne cedex, France}
\author{Serge Huant} \altaffiliation{Institut NEEL, CNRS and Universit\'e Grenoble Alpes, F-38000 Grenoble, France}
\author{Cyriaque Genet} \altaffiliation{ISIS, UMR 7006, CNRS-Universit\'e de Strasbourg, 67000 Strasbourg, France}
\author{Aur\'elien Drezet} \altaffiliation{Institut NEEL, CNRS and Universit\'e Grenoble Alpes, F-38000 Grenoble, France}\email{aurelien.drezet@neel.cnrs.fr}

\date{\today}

\begin{abstract}
We introduce a new paradigm: the chiral electromagnetic local density of states (LDOS) in a spiral plasmonic nanostructure. In both classical and quantum regimes, we reveal using scanning near-field optical microscopy (NSOM) in combination with spin analysis that a spiral cavity possesses spin-dependent local optical modes. We expect this work to lead to promising directions for future quantum plasmonic device development, highlighting the potentials of chirality in quantum information processing.
 
\end{abstract}

\pacs{73.20.Mf, 42.50.Ct, 07.79.Fc, 42.50.Tx, 42.25.Ja}
\maketitle

\section{Introduction}
Chiral metallic nanostructures \cite{Meinzer2014,Schaferling2012} generate great interests in the study and manipulation of light-matter interactions at the nanoscale and are expected to open new possibilities in applications ranging from biochemistry to highly integrated photonic circuits. For example, the ability of chiral plasmonic nanostructures to enhance and tailor optical near fields has been investigated in order to generate electromagnetic fields that can improve the chiroptical response of chiral molecules \cite{Valev2013,Hendry2010}. Spiral structures have received considerable attention \cite{Drezet2008,Gorodetski2013,Schnell2016} owing to their potential for controlling quantum source emission properties such as the spin and the directivity \cite{Rui2013,Gorodetski2009} as well as for their capability to generate surface plasmon polaritons (SPs) vortices with tailored orbital angular momentum (OAM) \cite{Gorodetski2013,Gorodetski2008}. The control of this additional degree of freedom is expected to find relevant applications in optical trapping \cite{Tsai2014} and optical communication systems \cite{Cai2012,Padgett2004}. Optical spin Hall effects arising from the spin-orbit interactions of incident photons with chiral plasmonic structures have also been reported recently \cite{Gorodetski2008,Gorodetski2012,Bliokh2015}. Moreover, with the aim of engineering light-matter coupling and to develop compact integrated devices involving quantum emitters, a full characterization of system's plasmonic properties in a chiral environment is required. This is the central motivation of this article and for this purpose we will consider the concept of local density of states (LDOS) for a chiral plasmonic environment.

The optical LDOS \cite{Carminati2015,Chicanne2002,ColasdesFrancs2001} provides an electromagnetic description of the system that is the optical modes available for emission for a given emitter position. What we will show is that the optical LDOS carries fundamental information about the chirality of the plasmonic system. We will demonstrate both theoretically and experimentally how to reveal this chirality hidden in the optical LDOS. This will in turn offer new tools for characterizing chiral plasmonic systems. So far, spiral cavities have been probed with photon scanning tunneling microscope in order to detect the SP near-field while it is illuminated with far-field methods \cite{Gorodetski2008}. Here, our study leverages on recent works which demonstrated that near-field illumination from a NSOM aperture tip can be used as a point-like light source to efficiently probe the optical LDOS~\cite{Chicanne2002,ColasdesFrancs2001}. This is the path taken in the present work for a proof of principle of chiral plasmonic LDOS measurement. More precisely, we show that the analysis of selected optical modes, i.e. partial optical LDOS, detected with high resolution in the momentum space reveals singular modal distribution which are dependent on both the spin and the chiral environment. 

Furthermore, it has been recently proven that optical LDOS mapping with a quantum fluorescent emitter \cite{Beams2013,Aigouy2014,Cao2015} is of significant interest. This is particularly true in the field of quantum information in which strong light-matter interactions between engineered nanostructures and quantum emitters offer new opportunities for the development of efficient and controlled single-photon sources \cite{Tame2013}. Therefore, in order to investigate the coupling of quantum emitters with a chiral plasmonic system we will in this letter implement chiro-optical LDOS measurements involving the recently developed quantum NSOM probe that uses nitrogen vacancy color centers as a quantum source of light~\cite{Cuche2009,Cuche2010}. Additionally, we show that the information on the dipole symmetry is crucial in the understanding and the control of the partial LDOS maps. All together the methodology presented in this work will provide a complete toolkit for probing the optical LDOS of chiral plasmonic systems both in the classical and quantum regime. A theoretical analysis based on coherent superposition of radiating electrical point-like dipoles distributed over the structure provides physical insights and faithfully reproduces the experimental NSOM images.

\begin{figure}[!htb]
\centering
\includegraphics[width=0.8\columnwidth]{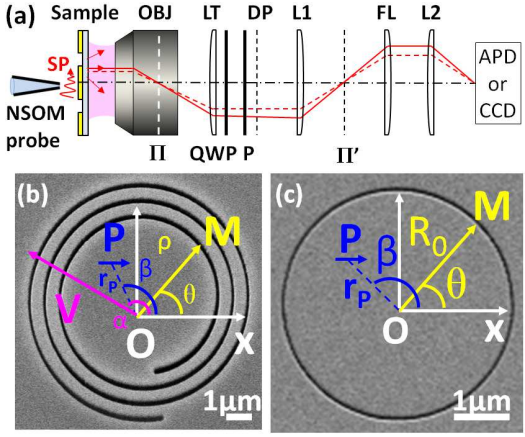}
\caption{(a) Sketch of the experimental setup: A linearly polarized laser beam is injected into the NSOM aperture probe or NV-based probe (not shown here). The later is brought in the near field of the sample to launch SPs. The scattered signal is collected by a microscope objective (OBJ). A removable Fourier lens (FL) enables to image the back focal plane ($\Pi$). Optical LDOS maps are obtained by raster scanning the sample while maintaining the probe in the near field of the sample. The signal is collected by an optical fiber connected to an APD. Wide-field images of both direct plane (DP) and back-focal plane ($\Pi'$) are captured by a CCD camera. A quarter wave plate (QWP) and polarizer (P) enable circular polarization analysis. Scanning electron microscope images of (a) a spiral and (b) circular plasmonic structure with geometrical parameters used in the theoretical study. Point $P$ represents the position of the tip with $\boldsymbol{r_p}=\boldsymbol{OP}:=[r_p,\beta]$ and $M$ the position of an electric dipole on the groove described by $\boldsymbol{OM}:=[\rho,\theta]$ for the spiral and $\boldsymbol{OM}:=[R_0,\theta]$ for the circular cavity. The pink arrow represents the vector $\boldsymbol{V}=k_{SP}\boldsymbol{r_p}+\rho\boldsymbol{k}$ for the spiral cavity defined as $\textbf{V}:=[V,\alpha]$}. \label{figure1}
\end{figure}

\section{Partial optical LDOS in a spiral plasmonic cavity}
\indent As explained in the motivations both classical and quantum NSOM probes are considered in this work. On one hand, the classical probe is made of a chemically etched single mode fiber, coated with 100 nm thick aluminum layer and with a typical 100 nm diameter aperture at the apex (see \cite{Chevalier2006,Novotny2006} for fabrication details). On the other hand, the quantum probe consists of nitrogen-vacancy (NV) centers hosted in a diamond nanocrystal, which have the advantage of featuring high photostability at room temperature. The nanodiamond is grafted at the apex of a bare optical tip as detailed in \cite{Cuche2009,Cuche2010,Berthel2015,Berthel2016}. A sketch of the experimental setup is shown in Fig.\ref{figure1}(a). A linearly polarized laser beam is injected into the fiber tip ($\lambda$=633 nm for the classical probe, $\lambda$=532 nm for the quantum probe) while it is brought in the near field of the sample surface to excite SPs. In order to study the optical LDOS chirality, we first consider a spiral cavity consisting of a gold film evaporated onto a glass plate in which grooves are milled via focused ion beam (See Fig.\ref{figure1}(b)). Then the LDOS maps are compared to that obtained for a ring-shaped plasmonic system with rotational symmetry (See Fig.\ref{figure1}(c)). 
\begin{figure}[!htb]
\centering
\includegraphics[width=0.8\columnwidth]{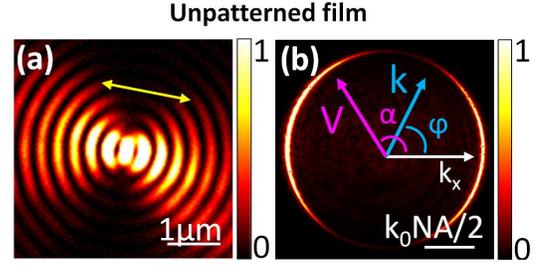}
\caption{SP excitation on a unpatterned 50 nm thick gold film by an aperture probe: CCD images of (a) the direct and (b) Fourier plane. In (a), a beam blocker is placed in the Fourier plane to remove the light directly transmitted through the sample. The yellow arrow indicates the polarization direction of the dipole. In (b), geometrical parameters used in the theoretical study: the blue arrow represents the wavevector defined as $\textbf{k}:=[k,\varphi]$, the pink arrow represents the vector $\boldsymbol{V}=k_{SP}\boldsymbol{r_p}+\rho\boldsymbol{k}$ for the spiral cavity defined as $\textbf{V}:=[V,\alpha]$}. \label{figure5}
\end{figure}
The scattered electromagnetic field is collected  using an oil immersion objective of high numerical aperture (NA) whose back focal plane can be imaged by a removable Fourier lens \cite{Berthel2015,Berthel2016}. For a given tip position, wide-field imaging of both direct and Fourier planes can be recorded via a CCD camera. Noteworthy, in case of a sufficiently thin metallic film (typically 50 nm thick), SPs leak through the metal, propagate into the glass substrate and can then be collected in the far field by leakage radiation microscopy (LRM) \cite{Hecht1996,Drezet2008Mater.Sci.Eng.B,Berthel2015,Drezet2013}. As they leak at a specific angle $\theta_{LR}$, corresponding to the wavevector matching condition $\Re[k_{SP}]=\frac{2\pi}{\lambda} n\sin{(\theta_{LR})}$ with $\Re[k_{SP}]$ standing for the real part of the SP in-plane wavevector, $n\simeq 1.5$ the glass optical index and $\lambda$ the optical wavelength, the SP waves appear in the Fourier plane on a ring of radius $\Re[k_{SP}]$. 

Since the partial LDOS is highly dependent on the dipole orientation, this LRM imaging technique is first implemented to characterize the field emitted by the classical probe (see Fig.\ref{figure5}(a),(b)). This is crucial in order to control the symmetry of the field emitted by the tip, thus the quality of our homemade NSOM probe. The sample surface is imaged as the tip is approached in the near field of an unpatterned 50 nm thick gold film. The analysis of the resulting SP field informs on the dipolar nature of the tip, which is characteristic of a SP emission launched by an electric in-plane dipole. These features, with two lobes in the direct plane (Fig.\ref{figure5}(a)) and two bright arcs in the Fourier plane (Fig.\ref{figure5}(b)), are expected for an aperture tip and have been studied in detail in \cite{Berthel2015}. The observed interference fringes in Fig.\ref{figure5}(a) are associated with an Airy diffraction pattern induced by the finite NA of the microscope objective~\cite{Hohenau,Berthel2015,Drezet2013}. Furthermore, in Fig.\ref{figure5}(b) we verify that the intensity of the uniform background ($k<k_{SP}$), which is due to the directly transmitted light from the tip, is sufficiently weak so we can neglect its contribution in the following experiments. 

\begin{figure*}[!htb]
\centering
\includegraphics[width=1.8\columnwidth]{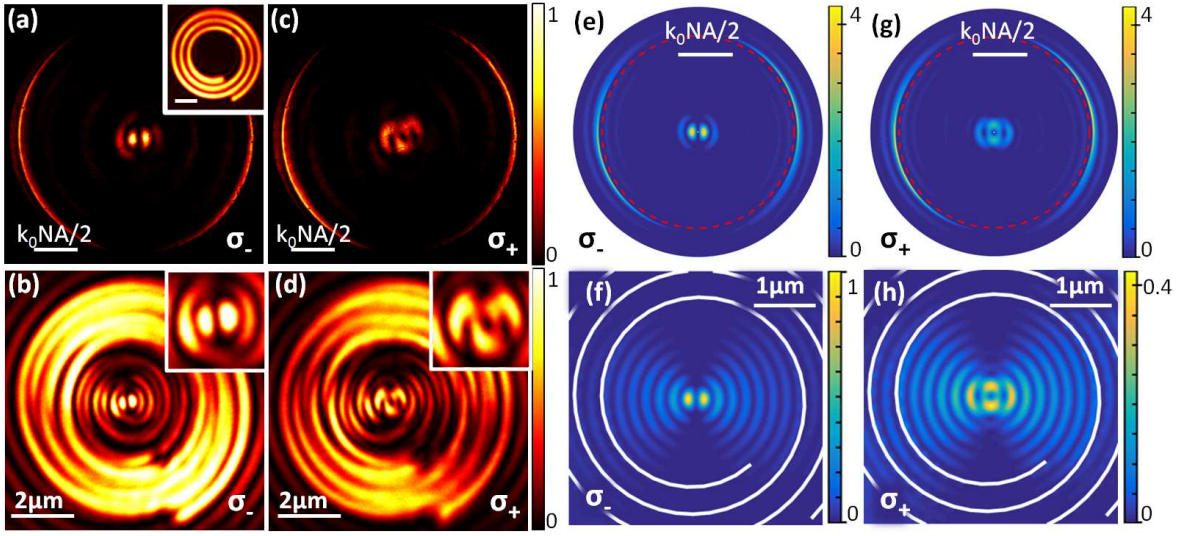}
\caption{SP excitation of a spiral structure by an aperture probe. (a), (c) Experimental Fourier plane images. Inset in (a): far-field image of the structure under study (Scale bar value: 2 $\mu$m). (b), (d) Experimental optical LDOS maps (Inset: zoom of the central area with the magnification 2.2x). (e-g) Theoretical simulations computed for a pure in-plane dipole: (e), (g) Fourier plane images,  (f), (h) optical LDOS maps. The red dashed circle indicates $k_0$. The white spiral indicates the position of the structure. Polarization analysis are performed as indicated by $\sigma_\pm$.} \label{figure3}
\end{figure*}

\indent Having identified the symmetry of the dipole, we begin with the study of the optical LDOS generated by a spiral plasmonic structure in the classical regime of excitation. The probe is placed at the origin of the spiral and approached at about 100 nm above the sample in order to launch the SPs. The spiral plasmonic structure is fabricated on a 50 nm thick gold layer (inset Fig.\ref{figure3}(a)) therefore allowing LRM imaging to verify the SPs. To reveal a break symmetry of the optical channels associated with a chiral environment, a quarter wave plate and a polarizer, placed after the microscope objective, allow the analysis of the modes in the left ($\sigma_+=+1$) and right ($\sigma_-=-1$) circular polarization basis. On the Fourier plane images recorded by the CCD camera (Fig.\ref{figure3}(a), (c)), we recover the two characteristic bright arcs at $k=\Re[k_{SP}]$ previously discussed as signature of the leakage of propagative SPs. Remarkably, the scattered radiation near the optical axis, i.e. $k\simeq 0$, features a distinct response according to whether the signal is analyzed in the $\sigma_{+}$ or $\sigma_{-}$ spin. This spin-dependent beam deflection is reminescent of an optical spin Hall effect \cite{Bliokh2008,Hosten2008,Yin2013}. 
The understanding of the imaged fields is analytically obtained by considering the signal as the result of a coherent superposition of elementary point sources uniformly distributed over the spiral groove. We recall that an Archimedean spiral is mathematically defined in polar basis ($\boldsymbol{\widehat{\rho}}$, $\boldsymbol{\widehat{\theta}}$) as $\rho(\theta)=\frac{\Lambda}{2\pi}\sigma_S\theta$ with $\theta\in[\theta_0=2\pi\rho_0/(\Lambda\sigma_S);\theta_0+2\pi n]$ where $n$ is the number of turns of the spiral, $\rho_0$=2 $\mu m$ is the starting distance of the spiral from the center, $\Lambda$ the pitch taken equal to the SP wavelength $\lambda_{SP}={2\pi}/k_{SP}$ to obtain resonant excitation of the SPs on the structured metal interface and $\sigma_S=1$ the geometrical charge (notation in Fig.\ref{figure1}(b)). The near-field illumination induces a SP field at point P that propagates outward until it decouples into the far field after scattering on the groove. Each point M of the structure then acts as a point-like electric dipole $\boldsymbol{\mu}e^{-i\omega t}$ radiating the electric field $\mathbf{E}(M)\propto (e^{ik_{SP}.\vert\mathbf{PM}\mid}/(\sqrt{\vert\mathbf{PM}\mid})(\boldsymbol{\mu}.\boldsymbol{\widehat{\rho}})\boldsymbol{\widehat{\rho}}$ \cite{Gorodetski2013,Drezet2008Mater.Sci.Eng.B,Hecht1996}.
As shown in the Appendix B, by integrating the fields of all the point-like sources over the spiral groove, we find that the total field is expressed as :
\begin{eqnarray}
\boldsymbol{E}(\boldsymbol{k},\boldsymbol{r_p}) \simeq\int^{\theta_0+2\pi n}_{\theta_0} \dfrac{e^{ik_{SP}\rho}}{\sqrt{\rho}} e^{-i\boldsymbol{V}.\boldsymbol{\widehat{\rho}}}(\boldsymbol{\mu}.\boldsymbol{\widehat{\rho}})\rho\boldsymbol{\widehat{\rho}}d\theta ,\\
\simeq \sum_n \sqrt{\rho_n}e^{ik_{SP}\rho_n}\oint e^{i\sigma_S\theta}e^{-iV\cos{(\alpha-\theta)}}(\boldsymbol{\mu}.\boldsymbol{\widehat{\rho}})\boldsymbol{\widehat{\rho}}d\theta, \label{eqn:eq1}
\end{eqnarray}
where $\boldsymbol{V}=k_{SP}\boldsymbol{r_p}-\rho\boldsymbol{k}$ and $\boldsymbol{V}:=[V,\alpha]$. As $ \rho_0 \gg \Lambda $, we made the approximation that a spiral making $n$ turns can be described as $n$ circles of radius $\rho_n=\rho_0+\Lambda\sigma_S n$.

In case of an excitation by an in-plane electric dipole, we have $\boldsymbol{\mu}=\mu\hat{\textbf{x}}$ hence $\boldsymbol{\mu}.\boldsymbol{\widehat{\rho}}=\mu \cos{(\theta)}$. Expressed in terms of the circular polarization basis $\boldsymbol{\widehat{\sigma}_\pm}=\frac{\mathbf{\widehat{x}}\pm i\mathbf{\widehat{y}}}{\sqrt{2}}$, Eq.\ref{eqn:eq1} then writes:
\begin{widetext}
\begin{equation}
\begin{split}
\boldsymbol{E}(\boldsymbol{k},\boldsymbol{r_p}) 
\propto \sum_n \sqrt{\rho_n}e^{ik_{SP}\rho_n}\lbrace \left((-i)^{\sigma_S}e^{i\sigma_S\alpha}J_{\sigma_S}(V)+(-i)^{\sigma_S-2\sigma_-}e^{i(\sigma_S-2\sigma_-)\alpha}J_{\sigma_S-2\sigma_-}(V)\right)\boldsymbol{\widehat{\sigma}}_-\\
+\left((-i)^{\sigma_S}e^{i\sigma_S\alpha}J_{\sigma_S}(V)+(-i)^{\sigma_S-2\sigma_+}e^{i(\sigma_S-2\sigma_+)\alpha}J_{\sigma_S-2\sigma_+}(V)\right)\boldsymbol{\widehat{\sigma}}_+\rbrace,
\end{split}
\label{eqn:eq3}
\end{equation} 
\end{widetext}
where $J_n$ denotes the Bessel function of $ n^{th} $ order. The intensity recorded in the Fourier plane for a given position of the tip (here $\boldsymbol{r_p}\approx0$) is proportional to $|\boldsymbol{E}(\boldsymbol{k},\boldsymbol{r_p}\approx0)|^2$ with $\boldsymbol{V}=\rho_n\boldsymbol{k}$. Projected on $\sigma_\pm$, the intensity is then given by:  \begin{equation}
\begin{split}
I_{\sigma_\pm}(\boldsymbol{k},\boldsymbol{r_p}\approx0)
\propto |\sum_n \sqrt{\rho_n}e^{ik_{SP}\rho_n}\lbrace(-i)^{\sigma_S}e^{i\sigma_S\varphi}J_{\sigma_S}(\rho_nk)\\
+(-i)^{\sigma_S-2\sigma_\pm}e^{i(\sigma_S-2\sigma_\pm)\varphi}J_{\sigma_S-2\sigma_\pm}(\rho_nk)\rbrace|^2,
\end{split}
\label{eqn:eq4}
\end{equation} 
with $\alpha=\varphi$ the angle between $\boldsymbol{k}$ and $\boldsymbol{k}_x$ as depicted in  (Fig.\ref{figure5} (b)). We demonstrate that the Fourier plane intensity distributions are thus defined as a combination of Bessel functions with orders that are directly related to the spin $\sigma_\pm$ and the chirality of the cavity $\sigma_S$. Our theoretical model emphasizes the crucial role of the chiral environment in the spin-orbit coupling which converts the SP field into free-space singular beams. Figs.\ref{figure3}(e), (g) are the corresponding theoretical images computed for a pure in-plane dipole excitation. It includes a more refined description of the optical system based on a rigorous modal expansion into TE and TM fields allowing us to take into account the imaging effects of the NSOM. For the sake of clarity, the present article does not present the full theoretical description of the NSOM that can be found in \cite{Berthel2015,Drezet2013}. The theoretical simulations faithfully reproduce the intensity distributions observed in the experimental images with two bright spots (Fig.\ref{figure3} (e)) for $\sigma_-$ and a four-lobe pattern (Fig.\ref{figure3}(g)) for $\sigma_+$.

Let us now turn on to the mapping of the optical LDOS of the chiral cavity. As detailed in \cite{Dereux2003}, scanning the near field of a structure with a tip that can be considered as an emitting point-like dipole $\boldsymbol{\mu}e^{i\omega t}$ allows to probe the optical modes available for emission at the position of the emitter. This can be achieved in NSOM working in illumination configuration by raster scanning the sample under the tip. For each position of the point-like source, the signal is collected by an optical fiber connected to an avalanche photodiode (APD), which records the signal to reconstruct a two dimensional image. Precisely, we demonstrate that a selective detection of a fraction of the radiative power, namely specific channels contained in the total LDOS, enables to highlight the chiral behavior of the optical modes. To this end, we recall that the total energy rate $P(\omega,\textbf{r}_p)$ associated with an electric dipole (accounting for radiative and non radiative exchanges with the environment) is related to the total LDOS as:
\begin{equation}
\rho(\omega,\textbf{r}_p)=\frac{3}{\pi\omega^2|\boldsymbol{\mu}|^2}P(\omega,\textbf{r}_p).
\label{eqn:eq5}
\end{equation} 
It is worth mentioning that in case of a two-level quantum emitter~\cite{Dung}, $P(\omega,\textbf{r}_p)=\hbar\omega\Gamma(\omega,\textbf{r}_p)$, with $\Gamma$ the full decay rate including the dissipation in the metal~\cite{Baffou2008}. One method to access to the full optical LDOS is then to measure the fluorescence and lifetime of emitters \cite{Cao2015,Krachmalnicoff}. However, the present study aims at investigating the modal spatial distribution associated with specific wavectors ($k\simeq 0$)  that cannot be readily identified by mapping the total LDOS. Indeed, one requires a highly resolved selection of the optical modes whereas very weak intensity distribution variation is expected from the integration over a wide number of miscellaneous modes, as confirmed in Ref.~\cite{submitted}. Here, we restrict the LDOS mapping to the spin-sensitive radiative channels by integrating, in the Fourier plane, a fraction of $P(\omega,\textbf{r}_p)$ over a wavevector area $\delta^2\textbf{k}\simeq(\omega/c)^2\pi r^2$ determined by the APD collection fiber (radius $r$=25 $\mu m$, 0.22NA)\cite{Berthel2016} which defines the resolution of the measurement (the non radiative contribution to the LDOS is therefore not recorded in our analysis). The latter is centered around the optical axis where prominent chiral behavior has been observed (Fig.\ref{figure3} (a), (c)). Accordingly, we define the partial optical LDOS integrated in $\delta^2\textbf{k}$ in terms of the recorded intensity  \cite{Carminati2015,Chicanne2002,ColasdesFrancs2001} such as 
\begin{equation}
\delta\rho^{\sigma_\pm}(\omega,\mathbf{k},\mathbf{r_p})=(3/\pi\mu^2\omega^2)|\boldsymbol{E}(\boldsymbol{k}\approx0,\boldsymbol{r_p}).\boldsymbol{\widehat{\sigma}}_\pm|^2\delta^2\textbf{k},
\label{eqn:eq6}
\end{equation} 
where the LDOS is projected onto the dipole orientation and circular polarization $\sigma_\pm$. From Eq.\ref{eqn:eq3}, we now have $\boldsymbol{V}=k_{SP}\boldsymbol{r_{p}}$ and $\alpha=\beta$ (Fig.\ref{figure1} (b)) , which leads:
\begin{equation}
\begin{split}
\delta\rho^{\sigma_\pm}(\omega,\mathbf{k}\approx 0,\mathbf{r_p})
\propto |\sum_n \sqrt{\rho_n}e^{ik_{SP}\rho_n}
\lbrace(-i)^{\sigma_S}e^{i\sigma_S\beta}
\\J_{\sigma_S}(k_{SP}r_p)+(-i)^{\sigma_S-2\sigma_\pm}e^{i(\sigma_S-2\sigma_\pm)\beta}J_{\sigma_S-2\sigma_\pm}(k_{SP}r_p)\rbrace|^2.
\end{split}
\label{eqn:eq7}
\end{equation} 
Strikingly, the partial optical LDOS exhibits a chiral behavior that is described by similar spin-sensitive Bessel function orders as the Fourier plane intensity (Eq.\ref{eqn:eq4}). It explains the strong similitude between the Fourier image patterns and the partial optical LDOS features depicted in Fig.\ref{figure3}(b) and (d) where a break of symmetry is also clearly visible. This important property can be revealed only if the modes at $k\approx 0$ are detected with a sufficiently high resolution, or ideally with a point-like detector. The two optical modes $\sigma_\pm$ show spin-dependent behavior in the momentum space which are specifically described by:
\begin{equation}
\begin{split}
\delta\rho^{\sigma_-}(\omega,\boldsymbol{k}\approx 0,\boldsymbol{r_p})
\propto |\sum_n \sqrt{\rho_n}e^{ik_{SP}\rho_n}|^2
\lbrace J^2_{1}(k_{SP}r_p)\\
+J^2_{3}(k_{SP}r_p)+J_{1}(k_{SP}r_p)J_{3}(k_{SP}r_p)\cos{(2\beta)}\rbrace ,
\end{split}
\label{eqn:eq8}
\end{equation} 
\begin{equation}
\begin{split}
\delta\rho^{\sigma_+}(\omega,\boldsymbol{k}\approx 0,\boldsymbol{r_p})\propto |\sum_n \sqrt{\rho_n}e^{ik_{SP}\rho_n}|^2\lbrace J^2_{1}(k_{SP}r_p)\\
+J^2_{-1}(k_{SP}r_p)+J_{1}(k_{SP}r_p)J_{-1}(k_{SP}r_p)\cos{(2\beta)}\rbrace.
\end{split}
\label{eqn:eq9}
\end{equation} 
 
\begin{figure}[!htb]
\centering
\includegraphics[width=\columnwidth]{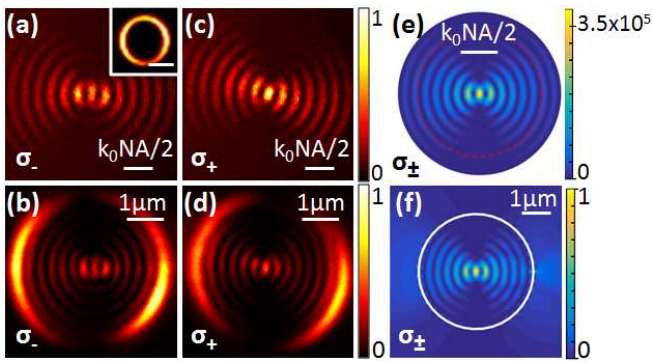}
\caption{SP excitation of a circular structure by an aperture probe. (a),(c) Experimental Fourier plane images and (b),(d) optical LDOS maps. Inset in (a): far-field image of the structure under study(Scale bar value: 2 $\mu$m). (e) Theoretical Fourier plane image and (f) optical LDOS map computed for a pure in-plane dipole. The red dashed circle indicates $k_0$. The white circle indicates the position of the structure. Circular polarization analysis are performed as indicated by $\sigma_\pm$.}\label{figure2}
\end{figure}

Close to the groove, we observe a spiral resulting from the light directly transmitted through the slit. It is worth emphasizing that the NSOM images can be related to the partial LDOS maps only in the central region of the structure away from the grooves, where one can consider that the probe field remains constant and is not notably modified by the reflected SPs\cite{Girard2005,submitted}. Eq.\ref{eqn:eq7} suggests that during the SP-light scattering process, the optical modes available for emission are constrained by an OAM selection rule, which allows only modal distributions with $(\sigma_{S})^{th}$ and $(\sigma_{S}-2\sigma_\pm)^{th}$ Bessel function orders. These singular modes with  distinct $\sigma_\pm$ channels are the manifestation of the chiral LDOS features possessed by the spiral and points out that the partial LDOS can be sculpted by adjusting the spiral geometrical charge $\sigma_{S}$. The very good agreement between the experimental LDOS maps and theoretical simulations computed for a pure in-plane dipole (Fig.\ref{figure3}(f),(h)) indicates a significant contribution of the dipole in-plane component of the NSOM tip. We ascribe the observed discrepancies to experimental errors (setup misalignment, polarizer uncertainties, tip position error) but also to a slight out-of-plane orientation of the dipole which was not accounted for in the theoretical model. This we think could potentially be used for probing small asymmetries in the tip due to chirality.

\begin{figure*}[!htb]
\centering
\includegraphics[width=1.8\columnwidth]{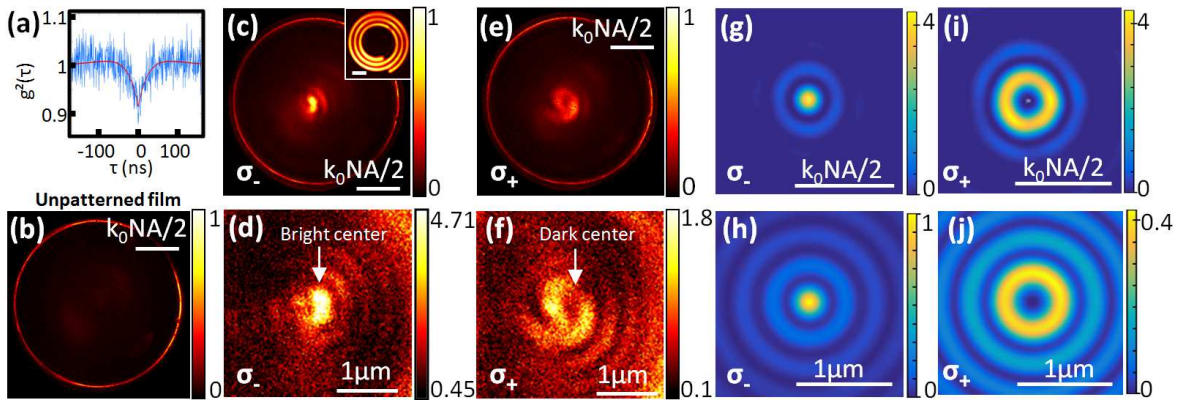}
\caption{(a) Normalized second-order correlation function $g^2$ as a function of the delay $\tau$ measured from the NV-based probe: experimental data in blue and fit in red curves. (b) LRM Fourier plane image of the SP excitation on an unpatterned 50 nm thick gold film by a quantum probe. (c-f) SP excitation by a quantum probe: (c), (e) Experimental LRM Fourier plane images. (d), (f) Experimental LDOS maps. Inset in (c): far-field image of the structure under study (Scale bar value: 2 $\mu$m). (g-j) Theoretical simulations computed for a pure out-of-plane dipole: (g), (i) Fourier planes, (h), (j) optical LDOS maps. Polarization analysis are performed as indicated by $\sigma_\pm$. } \label{figure4}
\end{figure*}

\indent To confirm the origin of the symmetry breaking observed in the optical modes produced by the spiral plasmonic system, we now investigate the LDOS supported by a cavity with an achiral symmetry. It consists of a $R_0$=2 $\mu$m radius circular groove (Fig.\ref{figure1}(c)) milled in a 200 nm thick gold layer. The SP are excited by the same classical probe as previously and placed at the origin of the ring. Contrary to what was found for the spiral cavity, we observe that both the Fourier plane images (Fig.\ref{figure2} (a), (c)) and partial LDOS maps (Fig.\ref{figure2} (b), (d)) exhibits  similar radiation pattern at the center regardless the helicity. Note that the thickness the film does not allow the SPs to leak and the fringes are associated with interferences between the scattered light from every opposite points of the groove leading to a constructive interference at the center, similarly to that occurring in a double slit experiment \cite{Wang2014}. Close to the groove position, we observe two side lobes resulting from the light directly transmitted through the slit. As it is expected, provided the rotational symmetry of the system, we derive an expression of the partial LDOS which does not feature spin-dependent behavior: 
\begin{equation}
\begin{split}
\delta\rho^{\sigma_\pm}(\omega,\boldsymbol{k}\approx 0,\boldsymbol{r_p})\propto J^2_0(k_{SP}r_p)+J^2_2(k_{SP}r_p)\\
-2J_0(k_{SP}r_p)J_2(k_{SP}r_p)\cos{(2\beta)}.
\end{split}
\label{eqn:eq10}
\end{equation}

For both $\sigma_\pm$, the modes are defined as a combination of $0^{th}$ and $2^{nd}$ order Bessel functions. We verify the agreement between the experimental and simulated partial LDOS maps displayed in Fig.\ref{figure2}(f). We thus retrieve that the circular polarized modes sustained by the ring cavity are indistinguishable. The chirality of the spiral cavity is confirmed to be at the origin of the spin-dependent optical modes. By revealing singular and chiral properties hidden in the optical LDOS, the presented methodology could be implemented for instance in the characterization of active plasmonic devices aiming at shaping and modulating the free space optical modes of an emitter in a chiral environment without resorting to bulky components. This is particularly relevant in the context of quantum integrated devices, which is the central motivation of the second part of our experiment: mapping of optical LDOS of the spiral cavity with an active NSOM probe based on a quantum source.

\section{Partial optical LDOS with a quantum probe}
\indent Whereas conventional fluorescent beads have been widely used for LDOS measurements, here we functionalized a NSOM probe with a nanodiamond containing NV color centers (typical diameter 80 nm) in order to extend the characterization method to the quantum regime of excitation, which is a critical issue in quantum nanophotonics (for probe functionalization details see \cite{Cuche2009,Cuche2010,Berthel2016}). In addition, owing to the photostability of this source, the NVs can, unlike the fluorescent beads, be excited in the saturation regime, while ensuring a constant dipole emission over time during the LDOS mapping. Compared with the previous LDOS mapping achieved with classical NSOM tip  (see Fig.~\ref{figure3}), the LDOS experiments with a quantum NSOM tip working in the saturation regime are therefore much more robust, i.e., even far away from the structure center.

Prior to carrying out the LDOS measurement, it is necessary to characterize the active NSOM probe. We first performed a second-order correlation function  measurement (see Fig.\ref{figure4}(a)). The observed antibunching dip infers on the quantum nature of the photon emission and we find that 10$\pm$1 NV centers comprise our optically active probe. Secondly, we deduce the symmetry of the dipole from the SP emission profile induced by the nanodiamond and measured by LRM on a bare 50 nm thick gold film. In Fig.\ref{figure4}(b), one observes a ring of uniform intensity which is characteristic of a radial SP emission. This can be ascribed to the random orientation of the NVs comprising in the nanodiamond. Determining the dipole orientation allows us to define accordingly the projected optical LDOS.

The quantum probe is then brought at the center of the spiral to excite SPs in the chiral cavity. Around $k \approx0$, we observe in the Fourier space (Fig.\ref{figure4}(c), (e)) an intense central spot when the signal is recorded in $\sigma_-$ whereas  in $\sigma_+$ we detect an annular pattern with a dark central spot which is characteristic of an optical vortex. These patterns are also observed in the optical LDOS maps (Fig.\ref{figure4}(d), (f)) that substantially differ from the response observed in Fig.\ref{figure3}(b), (d). This can be explained by the distinct emission profile of the NSOM tip used to probe the spiral. Indeed, as the SP coupling is higher in case of an excitation by an out-of-plane electric dipole than an in-plane one\cite{Berthel2015}, the contribution of resulting out-of-plane component of the randomly orientated NVs dominates. This assumption is verified by deriving the partial optical LDOS for a spiral excited by an out-of-plane dipole:
\begin{equation}
\begin{split}
\delta\rho^{\sigma_\pm}(\omega,\boldsymbol{k}\approx 0,\boldsymbol{r_p})\propto |\sum_n \mu\sqrt{\rho_n}e^{ik_{SP}\rho_n}|^2J^2_{\sigma_S+\sigma_\pm}(k_{SP}r_{p}).
\end{split}
\label{eqn:eq11}
\end{equation} 
As a result, we find that the optical LDOS features spin-dependent modes which are now related to $J^2_{0}(k_{SP}r_{p})$ for $\sigma_-$ and $J^2_{2}(k_{SP}r_{p})$ for $\sigma_-$. Precisely, the OAM selection rule achieved by the chiral structure in the spin-orbit interactions selectively allows circular polarized optical modes now described by $(\sigma_{S}+\sigma_\pm)^{th}$ order Bessel functions. In spite of the low number of photons emitted by the quantum probe, the optical LDOS mapping of the spiral cavity with the NV-based tip points out the potentials of our method in quantum communication where information can be encoded in polarization states and Bessel modes \cite{Padgett2004}. While the LDOS map simulations (see insets in Fig.\ref{figure4} (h), (j)) well agree with the experimental data, only the theoretical case of a purely vertical dipole has been considered. A more realistic model accounting for the in-plane contribution of the source could reproduce the observed divergences. Additionally, although we have measured the optical LDOS with a quantum probe containing a few number of NVs, the presented results suggests that our method can be extended to a single NV probe, for which the knowledge on its orientation will be critical for the understanding of the NSOM images. 

\section{Conclusions and perspectives}
\indent To conclude, the concept of chiral optical LDOS has been introduced. We have shown that using a NSOM in illumination configuration combined with resolved detection and circular polarization analysis allow to unfold the chirality of the optical modes supported by a spiral plasmonic structure. The optical partial LDOS maps were shown to exhibit, both in the classical and quantum regime, optical spin Hall effects and to follow an angular momentum selection rule. Specifically, our theoretical model analytically describes the recorded partial optical LDOS maps in terms of Bessel functions whose order can be directly related to the spin and the geometrical charge of the spiral.  We expect our method to bring an important contribution in the design and characterization of chiral plasmonic structures and envision its implementation in quantum plasmonic applications such as OAM control and beam shaping of light from quantum emitters\cite{Aouani2011,Jun2011,Lodhal}.
\section{Acknowledgments}
This work was supported by Agence Nationale de la Recherche (ANR), France, through the SINPHONIE (ANR-12-NANO-0019) and
PLACORE (ANR-13-BS10-0007) grants. Cyriaque Genet also thanks the ANR Equipex Union (ANR-10-EQPX-52-01). The Ph.D.
grant of A. Pham by the Minist\`ere de l'enseignement et la recherche, scientifique, and  of Q. Jiang by the R\'egion Rh\^one-Alpes is gratefully acknowledged. The authors thank J.-F. Motte and G. Julie, from NANOFAB facility at Institut Neel, for the optical tip manufacturing. They also thank T. Gacoin, from PMC laboratory at Ecole Polytechnique, and G. Dantelle, from Institut Neel, for providing the nanodiamonds used in this Article.

\appendix
\section{Circular plasmonic cavity}
This section provides the detailed derivations of our theoretical analysis which models the intensity distribution in the Fourier plane and the partial optical LDOS resulting from the excitation of the circular plasmonic structure by an in-plane and out-of-plane electric dipole.\\ 
The circular plasmonic structure of radius $R_0$ (See Fig.\ref{figure1} (c)) is excited by a NSOM probe located at point $P$, modeled as a point-like electric dipole $\boldsymbol{\mu}$. It gives rise to propagating SPs, which then radiated in the far field after scattering on the groove. Each point of the slit is considered to act as a dipole and to emit an electric field $\boldsymbol{\delta E}_{DP}$. In cylindrical coordinates $(\boldsymbol{\widehat{\rho}},\boldsymbol{\widehat{\theta}})$, it writes at point $M$ and in the direct plane as follows:
\begin{equation}
\begin{split}
\boldsymbol{\delta E}_{DP}(M)\propto \dfrac{e^{ik_{SP}\vert\boldsymbol{PM}\vert}}{\sqrt{\vert\boldsymbol{PM}\vert}}(\boldsymbol{\mu}.\boldsymbol{\widehat{\rho}})\boldsymbol{\widehat{\rho}}\\
\approx \mu \dfrac{e^{ik_{SP}R_0}}{\sqrt{R_0}}e^{-ik_{SP}\boldsymbol{r_p}\boldsymbol{\widehat{\rho}}}(\boldsymbol{\mu}.\boldsymbol{\widehat{\rho}})\boldsymbol{\widehat{\rho}},
\end{split} 
\label{eqA1}
\end{equation} 
with $k_{SP}$ the SP wavevector, $\textbf{OM}:=[R_0,\theta]$, $\textbf{r}_p=\textbf{OP}:=[r_p,\beta]$. The diffracted field $\boldsymbol{\delta E}_{FP}$ at point $M$ in the reciprocal space, for a given emission direction $\boldsymbol{k}$ can be expressed as $\boldsymbol{\delta E}_{FP}(\boldsymbol{k},\boldsymbol{r_p}) \propto f(k) \boldsymbol{\delta E}_{DP}(M)e^{-i\boldsymbol{k}.R_0\boldsymbol{\widehat{\rho}}}$ with $f(k)$ the point spread function of the system. For $k<NA$, $f(k)$ is assumed to be constant as verified in the experimental Fourier plane images (See Fig.\ref{figure5}(b), \ref{figure4}(b)). We assume that the probe is placed at the center of the ring such as $\boldsymbol{r_p}\ll R_0$. By integrating the field radiated over the entire plasmonic ring, we find: 

\begin{equation}
\begin{split}
\boldsymbol{E}_{FP}(\boldsymbol{k},\boldsymbol{r_p})\approx \oint \boldsymbol{\delta E}_{DP}(M)e^{-i\boldsymbol{k}.R_0\boldsymbol{\widehat{\rho}}}R_0d\theta
\\
=\sqrt{R_0}e^{ik_{SP}R_0} \oint e^{-i\boldsymbol{V}.\boldsymbol{\widehat{\rho}}}(\boldsymbol{\mu}.\boldsymbol{\widehat{\rho}})\boldsymbol{\widehat{\rho}} d\theta\\
=\sqrt{R_0}e^{ik_{SP}R_0}\oint e^{-iV\cos{(\alpha-\theta)}}(\boldsymbol{\mu}.\boldsymbol{\widehat{\rho}})\boldsymbol{\widehat{\rho}} d\theta,
\end{split}
\label{eqA2}
\end{equation}  
with $\boldsymbol{V}=k_{SP}\boldsymbol{r_p}+R_0\boldsymbol{k}$ and $\boldsymbol{V}:=[V,\alpha]$.

\subsection{In-plane electric dipole}
In case of an excitation by an in-plane electric dipole, we have $\boldsymbol{\mu}=\mu\hat{\textbf{x}}$ hence $\boldsymbol{\mu}.\boldsymbol{\widehat{\rho}}=\mu \cos{(\theta)}$. The projection on the circular polarization states $\boldsymbol{\widehat{\sigma}_\pm}=\frac{\mathbf{\widehat{x}}\pm i\mathbf{\widehat{y}}}{\sqrt{2}}$, yields: 
\begin{equation}
\begin{split}
\boldsymbol{E}_{FP}(\boldsymbol{k},\boldsymbol{r_p}) \simeq \mu e^{ik_{SP}R_0}\sqrt{R_0} \oint e^{-iV\cos{(\alpha-\theta)}}\\
\lbrace (1+e^{2i\theta)})\boldsymbol{\widehat{\sigma}}_-
+(1+e^{-2i\theta})\boldsymbol{\widehat{\sigma}}_+ \rbrace d\theta\\
\propto \mu e^{ik_{SP}R_0}\sqrt{R_0} \lbrace (J_0(V)-J_2(V)e^{i2\alpha})\boldsymbol{{\widehat{\sigma}}}_-\\
+(J_0(V)-J_2(V)e^{-i2\alpha})\boldsymbol{{\widehat{\sigma}}}_+\rbrace,
\end{split}
\label{eqA3}
\end{equation} 
where $ J_n $ is the $ n^{th} $ order Bessel function. 

On one hand, $|\boldsymbol{E}_{FP}(\boldsymbol{k},\boldsymbol{r_p}\approx0)|^2$ is proportional to the intensity recorded in the Fourier plane for a position of the tip $\boldsymbol{r_p}\approx0$, therefore $\boldsymbol{V}=R_0\boldsymbol{k}$. Projected on $\sigma_\pm$, the intensity is given by: 
\begin{equation}
\begin{split}
I_{\sigma_\pm}(\boldsymbol{k},\boldsymbol{r_p}\approx 0)\propto J^2_0(R_0k)+J^2_2(R_0k)\\
-2J_0(R_0k)J_2(R_0k)\cos{(2\varphi)},
\end{split}
\label{eqA4}
\end{equation} 
with $\alpha=\varphi$ the angle between $\boldsymbol{k}$ and $\boldsymbol{k}_x$ (See Fig.\ref{figure5}(b)).
On the other hand, the partial LDOS map corresponding to the experimental configuration of the article ($k\approx 0$) is obtained from $|\boldsymbol{E}_{FP}(\boldsymbol{k}\approx0,\boldsymbol{r_p})|^2$, therefore $\boldsymbol{V}=k_{SP}\boldsymbol{r_{p}}$:
\begin{equation}
\begin{split}
\delta\rho^{\sigma_\pm}(\boldsymbol{k}\approx 0,\boldsymbol{r_p})\propto J^2_0(k_{SP}r_p)+J^2_2(k_{SP}r_p)\\
-2J_0(k_{SP}r_p)J_2(k_{SP}r_p)\cos{(2\beta)},
\end{split}
\label{eqA5}
\end{equation}
with $\alpha=\beta$ the angle between $\boldsymbol{r_p}$ and $\boldsymbol{\widehat{x}}$ (See Fig.\ref{figure1}(b)).
\subsection{Out-of-plane electric dipole}
In case of an excitation by an out-of-plane electric dipole, we have $\boldsymbol{\mu}=\mu\boldsymbol{\hat{\rho}}$ hence $ \boldsymbol{\mu}.\boldsymbol{\widehat{\rho}}=\mu$. Eq.\ref{eqA2} becomes:
\begin{equation}
\begin{split}
\boldsymbol{E}_{FP}(\boldsymbol{k},\boldsymbol{r_p})
\propto\mu\sqrt{R_0}e^{ik_{SP}R_0}\oint e^{-iV\cos{(\alpha-\theta)}}\boldsymbol{\widehat{\rho}} d\theta\\
\propto\mu\sqrt{R_0}e^{ik_{SP}R_0}\lbrace e^{i\alpha}J_1(V)\boldsymbol{{\widehat{\sigma}}}_-+e^{-i\alpha}J_1(V)\boldsymbol{{\widehat{\sigma}}}_+\rbrace.
\end{split}
\label{eqA6}
\end{equation}
The intensity recorded in the Fourier plane for a position of the tip $\boldsymbol{r_p}\approx0$ is proportional to $|\boldsymbol{E}_{FP}(\boldsymbol{k},\boldsymbol{r_p}\approx0)|^2$, therefore $\boldsymbol{V}=R_0\boldsymbol{k}$. Projected on $\sigma_\pm$, the intensity is given by: 
\begin{equation}
I_{\sigma_\pm}(\boldsymbol{k},\boldsymbol{r_p}\approx0)\propto J^2_1(R_0k).
\label{eqA7}
\end{equation} 
The partial LDOS map corresponding to the experimental configuration of the article ($k\approx 0$) is obtained from $|\boldsymbol{E}_{FP}(\boldsymbol{k}\approx0,\boldsymbol{r_p})|^2$, therefore $\boldsymbol{V}=k_{SP}\boldsymbol{r_{p}}$. Projected on $\sigma_\pm$, it is given by: 
\begin{equation}
\delta\rho^{\sigma_\pm}(\boldsymbol{k}\approx 0,\boldsymbol{r_p})\propto J^2_1(k_{SP}r_p).
\label{eqA8}
\end{equation} 
\section{Spiral plasmonic cavity}
The spiral equation in cylindrical coordinates writes: $\rho=\frac{\Lambda}{2\pi}\sigma_S\theta$ with $\Lambda$ the pitch taken equal to $\lambda_{SP}$, $\sigma_S$ the geometrical charge, $\theta\in[\theta_0=2\pi\rho_0/(\Lambda\sigma_S);\theta_0+2\pi n]$, $n$ the number of turns, $\rho_0$ the starting distance of the spiral from the center. As previously, the spiral plasmonic structure (See Fig.\ref{figure1} (b)) is excited by a probe located at point $P$, modeled as a point-like electric dipole. The integration over the structure of the field emitted by each point M of the groove leads:
\begin{equation*}
\begin{split}
\boldsymbol{E}_{FP}(\boldsymbol{k},\boldsymbol{r_p}) \propto \int^{\theta_0+2\pi n}_{\theta_0} \dfrac{e^{ik_{SP}\vert\boldsymbol{PM}\vert}}{\sqrt{\vert\boldsymbol{PM}\vert}}e^{-i\boldsymbol{k}.\rho\boldsymbol{\widehat{\rho}}}(\boldsymbol{\mu}.\boldsymbol{\widehat{\rho}})(\boldsymbol{\widehat{\rho}}.\boldsymbol{\widehat{n}})\boldsymbol{\widehat{n}} dl\\
\end{split}
\end{equation*} 
\begin{equation*}
\begin{split}
\approx \int^{\theta_0+2\pi n}_{\theta_0} \dfrac{e^{ik_{SP}\rho}}{\sqrt{\rho}} e^{-i(k_{SP}\boldsymbol{r_p}+\rho\boldsymbol{k})\boldsymbol{\widehat{\rho}}}(\boldsymbol{\mu}.\boldsymbol{\widehat{\rho}})\rho\boldsymbol{\widehat{\rho}}d\theta,
\end{split}
\end{equation*}
\begin{equation}
\begin{split}
=\int^{\theta_0+2\pi n}_{\theta_0} \dfrac{e^{ik_{SP}\rho}}{\sqrt{\rho}} e^{-iV\cos{(\alpha-\theta)}}(\boldsymbol{\mu}.\boldsymbol{\widehat{\rho}})\rho\boldsymbol{\widehat{\rho}}d\theta,
\end{split}
\label{eqB1}
\end{equation} 
where we made the approximation $\theta \gg 1$, $\boldsymbol{\widehat{n}}$ is a unit vector normal to the spiral, $dl$ the length element, $\boldsymbol{V}=k_{SP}\boldsymbol{r_p}+\rho\boldsymbol{k}$ and $\boldsymbol{V}:=[V,\alpha]$. As $ \rho_0 \gg \Lambda $, we make the approximation that a spiral making $n$ turns can be described as $n$ circles of radius \textbf{$\rho_n=\rho_0+\Lambda\sigma_S n$}. Eq.\ref{eqB1} becomes:
\begin{equation}
\begin{split}
\boldsymbol{E}_{FP}(\boldsymbol{k},\boldsymbol{r_p}) \simeq \sum_n \oint \sqrt{\rho_n+\frac{\Lambda\sigma_S}{2\pi}\theta}e^{ik_{SP}(\rho_n+\frac{\Lambda\sigma_S}{2\pi}\theta)} \\
e^{-iV\cos{(\alpha-\theta)}}(\boldsymbol{\mu}.\boldsymbol{\widehat{\rho}})\boldsymbol{\widehat{\rho}}d\theta\\
\approx \sum_n \sqrt{\rho_n}e^{ik_{SP}\rho_n}\oint e^{i\sigma_S\theta}e^{-iV\cos{(\alpha-\theta)}}(\boldsymbol{\mu}.\boldsymbol{\widehat{\rho}})\boldsymbol{\widehat{\rho}}d\theta,
\end{split}
\label{eqB2}
\end{equation}
where we made the approximation $\rho_n \gg \Lambda$.
\subsection{In-plane electric dipole}
In case of an excitation by an in-plane electric dipole, we have $\boldsymbol{\mu}=\mu\hat{\textbf{x}}$ hence $\boldsymbol{\mu}.\boldsymbol{\widehat{\rho}}=\mu \cos{(\theta)}$. Therefore: 
\begin{widetext}
\begin{equation}
\begin{split}
\boldsymbol{E}_{FP}(\boldsymbol{k},\boldsymbol{r_p}) \simeq \sum_n \sqrt{\rho_n}e^{ik_{SP}\rho_n}\oint e^{i\sigma_S\theta}e^{-iV\cos{(\alpha-\theta)}}\frac{e^{i\theta}+e^{-i\theta}}{2} \frac{e^{i\theta}\boldsymbol{\widehat{\sigma}}_-+e^{-i\theta}\boldsymbol{\widehat{\sigma}}_+}{\sqrt{2}}d\theta\\
\propto \sum_n \sqrt{\rho_n}e^{ik_{SP}\rho_n}\lbrace ((-i)^{\sigma_S}e^{i\sigma_S\alpha}J_{\sigma_S}(V)
+(-i)^{\sigma_S-2\sigma_-}e^{i(\sigma_S-2\sigma_-)\alpha}J_{\sigma_S-2\sigma_-}(V))\boldsymbol{\widehat{\sigma}}_-\\
+((-i)^{\sigma_S}e^{i\sigma_S\alpha}J_{\sigma_S}(V)+
(-i)^{\sigma_S-2\sigma_+}e^{i(\sigma_S-2\sigma_+)\alpha}J_{\sigma_S-2\sigma_+}(V))\boldsymbol{\widehat{\sigma}}_+\rbrace ,
\end{split}
\label{eqB3}
\end{equation} 
\end{widetext}
where $ J_n $ is the $ n^{th} $ order Bessel function. 
On one hand, $|\boldsymbol{E}_{FP}(\boldsymbol{k},\boldsymbol{r_p}\approx0)|^2$ is proportional to the intensity recorded in the Fourier plane for a position of the tip $\boldsymbol{r_p}\approx0$, therefore $\boldsymbol{V}=\rho_n\boldsymbol{k}$. Projected on $\sigma_\pm$, the intensity is given by: 
\begin{equation}
\begin{split}
I_{\sigma_-}(\boldsymbol{k},\boldsymbol{r_p}\approx0)
\propto |\sum_n \sqrt{\rho_n}e^{ik_{SP}\rho_n}\lbrace(-i)^{\sigma_S}e^{i\sigma_S\varphi}J_{\sigma_S}(\rho_nk)\\
+(-i)^{\sigma_S-2\sigma_-}e^{i(\sigma_S-2\sigma_-)\varphi}J_{\sigma_S-2\sigma_-}(\rho_nk)\rbrace|^2 ,\\
I_{\sigma_+}(\boldsymbol{k},\boldsymbol{r_p}\approx0)
\propto |\sum_n \sqrt{\rho_n}e^{ik_{SP}\rho_n}\lbrace(-i)^{\sigma_S}e^{i\sigma_S\varphi}J_{\sigma_S}(\rho_nk)+\\
(-i)^{\sigma_S-2\sigma_+}e^{i(\sigma_S-2\sigma_+)\varphi}J_{\sigma_S-2\sigma_+}(\rho_nk)\rbrace|^2,
\end{split}
\label{eqB4}
\end{equation} 
with $\alpha=\varphi$ the angle between $\boldsymbol{k}$ and $\boldsymbol{k}_x$.

On the other hand, the partial LDOS map corresponding to the experimental configuration of the article ($k\approx 0$) is obtained from $|\boldsymbol{E}_{FP}(\boldsymbol{k}\approx0,\boldsymbol{r_p})|^2$, therefore $\boldsymbol{V}=k_{SP}\boldsymbol{r_{p}}$:
\begin{equation}
\begin{split}
\delta\rho^{\sigma_-}(\boldsymbol{k}\approx 0,\boldsymbol{r_p})\propto |\sum_n \sqrt{\rho_n}e^{ik_{SP}\rho_n}\lbrace(-i)^{\sigma_S}e^{i\sigma_S\beta}\\
J_{\sigma_S}(k_{SP}r_p)+
(-i)^{\sigma_S-2\sigma_-}e^{i(\sigma_S-2\sigma_-)\beta}J_{\sigma_S-2\sigma_-}(k_{SP}r_p)\rbrace|^2\\
\propto |\sum_n \sqrt{\rho_n}e^{ik_{SP}\rho_n}|^2\lbrace J^2_{\sigma_S}(k_{SP}r_p)+J^2_{\sigma_S-2\sigma_-}(k_{SP}r_p)\\
+J_{\sigma_S}(k_{SP}r_p)J_{\sigma_S-2\sigma_-}(k_{SP}r_p)\cos{(2\beta)}\rbrace ,\\
\end{split}
\label{eqB5}
\end{equation} 

\begin{equation}
\begin{split}
\delta\rho^{\sigma_+}(\boldsymbol{k}\approx 0,\boldsymbol{r_p})
\propto |\sum_n \sqrt{\rho_n}e^{ik_{SP}\rho_n}\lbrace(-i)^{\sigma_S}e^{i\sigma_S\beta}\\
J_{\sigma_S}(k_{SP}r_p)+
(-i)^{\sigma_S-2\sigma_+}e^{i(\sigma_S-2\sigma_+)\beta}J_{\sigma_S-2\sigma_+}(k_{SP}r_p)\rbrace|^2\\
\propto |\sum_n \sqrt{\rho_n}e^{ik_{SP}\rho_n}|^2\lbrace J^2_{\sigma_S}(k_{SP}r_p)+J^2_{\sigma_S-2\sigma_+}(k_{SP}r_p)\\
+J_{\sigma_S}(k_{SP}r_p)J_{\sigma_S-2\sigma_+}(k_{SP}r_p)\cos{(2\beta)}\rbrace ,
\end{split}
\label{eqB6}
\end{equation} 
with $\alpha=\beta$ the angle between $\boldsymbol{r_p}$ and $\boldsymbol{\widehat{x}}$.
\subsection{Out-of-plane electric dipole}
In case of an excitation by an out-of-plane electric dipole, we have $\boldsymbol{\mu}=\mu\boldsymbol{\hat{\rho}}$ hence $ \boldsymbol{\mu}.\boldsymbol{\widehat{\rho}}=\mu$. 
\begin{equation*}
\begin{split}
\boldsymbol{E}_{FP}(\boldsymbol{k},\boldsymbol{r_p}) \simeq  \sum_n \mu\sqrt{\rho_n}e^{ik_{SP}\rho_n}\oint e^{i\sigma_S\theta}e^{-iV\cos{(\alpha-\theta)}}\boldsymbol{\widehat{\rho}}d\theta\\
\propto \sum_n \mu\sqrt{\rho_n}e^{ik_{SP}\rho_n}\lbrace e^{i\alpha}\oint e^{-iVcos\theta}e^{i(\sigma_S+\sigma_-)\theta}\boldsymbol{\widehat{\sigma}}_-d\theta\\
+e^{-i\alpha}\oint e^{-iVcos\theta}e^{i(\sigma_S+\sigma_+)\theta}\boldsymbol{\widehat{\sigma}}_+d\theta\rbrace\\
\end{split}
\end{equation*}
\begin{equation}
\begin{split}
\propto \sum_n \mu\sqrt{\rho_n}e^{ik_{SP}\rho_n}\lbrace (-i)^{\sigma_S+\sigma_-}e^{i\alpha}J_{\sigma_S+\sigma_-}(V)\boldsymbol{\widehat{\sigma}}_-\\
+ (-i)^{\sigma_S+\sigma_+}e^{i\alpha}J_{\sigma_S+\sigma_+}(V)\boldsymbol{\widehat{\sigma}}_+\rbrace ,\\
\label{eqB7}
\end{split}
\end{equation}
On one hand, $|\boldsymbol{E}_{FP}(\boldsymbol{k},\boldsymbol{r_p}\approx0)|^2$ is proportional to the intensity recorded in the Fourier plane for a position of the tip $\boldsymbol{r_p}\approx0$, therefore $\boldsymbol{V}=\rho_n\boldsymbol{k}$. Projected on $\sigma_\pm$, the intensity is given by: 
\begin{equation}
\begin{split}
I_{\sigma_-}(\boldsymbol{k},\boldsymbol{r_p}\approx0)\propto |\sum_n \mu\sqrt{\rho_n}e^{ik_{SP}\rho_n}\\
\lbrace (-i)^{\sigma_S+\sigma_-}e^{i\varphi}J_{\sigma_S+\sigma_-}(\rho_nk)\rbrace|^2 ,\\
\end{split}
\label{eqB8}
\end{equation} 
\begin{equation}
\begin{split}
I_{\sigma_+}(\boldsymbol{k},\boldsymbol{r_p}\approx0)\propto |\sum_n \mu\sqrt{\rho_n}e^{ik_{SP}\rho_n}\\
\lbrace (-i)^{\sigma_S+\sigma_+}e^{i\varphi}J_{\sigma_S+\sigma_+}(\rho_nk)\rbrace|^2 ,\\
\end{split}
\label{eqB9}
\end{equation} 
with $\alpha=\varphi$ the angle between $\boldsymbol{k}$ and $\boldsymbol{k}_x$.

On the other hand, the partial LDOS map corresponding to the experimental configuration of the article ($k\approx 0$) is obtained from $|\boldsymbol{E}_{FP}(\boldsymbol{k}\approx0,\boldsymbol{r_p})|^2$, therefore $\boldsymbol{V}=k_{SP}\boldsymbol{r_{p}}$:
\begin{equation}
\begin{split}
\delta\rho^{\sigma_-}(\boldsymbol{k}\approx 0,\boldsymbol{r_p})\propto |\sum_n \mu\sqrt{\rho_n}e^{ik_{SP}\rho_n}\lbrace (-i)^{\sigma_S+\sigma_-}e^{i\beta}\\
J_{\sigma_S+\sigma_-}(k_{SP}r_{p})\rbrace|^2\\
\propto |\sum_n \mu\sqrt{\rho_n}e^{ik_{SP}\rho_n}|^2J^2_{\sigma_S+\sigma_-}(k_{SP}r_{p}),\\
\delta\rho^{\sigma_+}(\boldsymbol{k}\approx 0,\boldsymbol{r_p}) \propto |\sum_n \mu\sqrt{\rho_n}e^{ik_{SP}\rho_n}\lbrace (-i)^{\sigma_S+\sigma_+}e^{i\beta}\\
J_{\sigma_S+\sigma_+}(k_{SP}r_{p})\rbrace|^2\\
\propto |\sum_n \mu\sqrt{\rho_n}e^{ik_{SP}\rho_n}|^2J^2_{\sigma_S+\sigma_+}(k_{SP}r_{p}),
\end{split}
\label{eq10}
\end{equation} 
with $\alpha=\beta$ the angle between $\boldsymbol{r_p}$ and $\boldsymbol{\widehat{x}}$.


\begin{thebibliography}{99}
\bibitem{Meinzer2014}
N. Meinzer, W.L. Barnes, and I.R. Hooper, \href{http://www.nature.com/nphoton/journal/v8/n12/full/nphoton.2014.247.html}{Nature Photonics \textbf{8}, 889-898 (2014)}.

\bibitem{Schaferling2012}
M. Schaferling, D. Dregely, M. Hentschel, and  H. Giessen, \href{http://journals.aps.org/prx/abstract/10.1103/PhysRevX.2.031010}{Phys. Rev. X \textbf{2}, 031010 (2012)}.

\bibitem{Valev2013}
V.K. Valev, J.J. Baumberg, C. Sibilia, and T. Verbiest,  
\href{http://onlinelibrary.wiley.com/doi/10.1002/adma.201205178/full}{Adv. Mater. \textbf{25}, 2517–2534 (2013)}.

\bibitem{Hendry2010}
E. Hendry, T. Carpy, J. Johnston, M. Popland, R. V. Mikhaylovskiy, A. J. Lapthorn, S. M. Kelly, L. D. Barron, N. Gadegaard, and M. Kadodwala, \href{http://www.nature.com/nnano/journal/v5/n11/abs/nnano.2010.209.html} {Nature Nanotech. \textbf{5}, 783–787 (2010)}.

\bibitem{Drezet2008}
A. Drezet, C. Genet, J.-Y. Laluet, and T. W. Ebbesen \href{https://www.osapublishing.org/oe/abstract.cfm?uri=oe-16-17-12559}{Optics Express \textbf{6} 12559-12570 (2008)}.

\bibitem{Gorodetski2013}
Y. Gorodetski, A. Drezet, C. Genet, and T.W. Ebbesen, \href{http://journals.aps.org/prl/abstract/10.1103/PhysRevLett.110.203906}{Phys. Rev. Lett. \textbf{110}, 203906 (2013)}.

\bibitem{Schnell2016}
M. Schnell, P. Sarriugarte, T. Neuman, A. B. Khanikaev, G. Shvets, J. Aizpurua, and R. Hillenbrand, \href{http://pubs.acs.org/doi/abs/10.1021/acs.nanolett.5b04416}{Nano Lett. \textbf{16}, 663–670 (2016)}.

\bibitem{Rui2013}
G. Rui, D.C. Abeysinghe, R.L. Nelson, and Q. Zhan, \href{http://www.nature.com/articles/srep02237}{Scientific Reports \textbf{3}, 2237 (2013)}.

\bibitem{Gorodetski2009}
Y. Gorodetski, A. Niv, N. Shitrit, V. Kleiner, and E. Hasman,  \href{http://pubs.acs.org/doi/abs/10.1021/nl901437d}{Nano Lett. \textbf{9}, 3016 (2009)}.

\bibitem{Gorodetski2008}
Y. Gorodetski, A. Niv, V. Kleiner, and E. Hasman, \href{http://journals.aps.org/prl/abstract/10.1103/PhysRevLett.101.043903}{Phys. Rev. Lett. \textbf{101}, 043903 (2008)}.

\bibitem{Tsai2014}
W.Y. Tsai, J.S. Huang, and C.B. Huang, \href{http://pubs.acs.org/doi/abs/10.1021/nl403608a}{Nano Lett. \textbf{14}, 547 (2014)}.

\bibitem{Cai2012}
X. Cai, J. Wang, M.J. Strain, B. Johnson-Morris, J. Zhu, M. Sorel, J.L. O’Brien, M.G. Thompson, and S. Yu, \href{http://science.sciencemag.org/content/338/6105/363}{Science \textbf{338}, 363 (2012)}.

\bibitem{Padgett2004}
M. Padgett, J. Courtial, and L. Allen,  \href{http://scitation.aip.org/content/aip/magazine/physicstoday/article/57/5/10.1063/1.1768672}{Physics Today \textbf{57}, 35-40 (2004)}.

\bibitem{Gorodetski2012}
Y. Gorodetski,  K. Y. Bliokh, B. Stein, C. Genet, N. Shitrit, V. Kleiner, E. Hasman, and T. W. Ebbesen, \href{http://journals.aps.org/prl/abstract/10.1103/PhysRevLett.109.013901}{Phys. Rev. Lett. \textbf{109}, 013901 (2012)}. 

\bibitem{Bliokh2015}
K. Y. Bliokh, F. J. Rodr{\'i}guez-Fortu{\~n}o, F. Nori and A.V. Zayats, \href{http://www.nature.com/nphoton/journal/v9/n12/full/nphoton.2015.201.html}{ Nature Photonics, \textbf{9(12)}, 796-808 (2015)}.

\bibitem{Carminati2015}
R. Carminati, A. Caz\'e, D. Cao, F. Peragut, V. Krachmalnicoff,  R. Pierrat, and Y. De Wilde, \href{http://www.sciencedirect.com/science/article/pii/S0167572914000338}{Surface Science Reports \textbf{70}, 1–41 (2015)}.

\bibitem{Chicanne2002}
C. Chicanne, T. David, R. Quidant, J.C. Weeber, Y. Lacroute, E. Bourillot, A. Dereux, G. Colas des Francs, and C. Girard, \href{http://journals.aps.org/prl/abstract/10.1103/PhysRevLett.88.097402}{Phys. Rev. Lett. \textbf{88}, 097402 (2002)}.

\bibitem{ColasdesFrancs2001}
G. Colas des Francs, C. Girard, J.C. Weeber, C. Chicane, T. David,  A. Dereux, and D. Peyrade, \href{http://journals.aps.org/prl/abstract/10.1103/PhysRevLett.86.4950}{Phys. Rev. Lett. \textbf{86}, 4950 (2001)}.

\bibitem{Beams2013}
R. Beams, D. Smith, T.W. Johnson, S.-H .Oh, L. Novotny, and A.N.Vamivakas, \href{http://pubs.acs.org/doi/abs/10.1021/nl401791v}{Nano Lett. \textbf{13}, 3807−3811 (2013)}.

\bibitem{Aigouy2014} 
L. Aigouy, A. Caz\'e, P. Gredin, M. Mortier, and R. Carminati,
\href{http://journals.aps.org/prl/abstract/10.1103/PhysRevLett.113.076101}{Phys. Rev. Lett. \textbf{113}, 076101 (2014)}.

\bibitem{Cao2015}
D. Cao, A. Caz\'e, M. Calabrese, R. Pierrat, N. Bardou, S. Collin, R. Carminati, V. Krachmalnicoff, and Y. De Wilde, \href{http://pubs.acs.org/doi/abs/10.1021/ph500431g}{ACS Photon. \textbf{2}, 189-193 (2015)}.

\bibitem{Tame2013}
M.S. Tame,	K.R. McEnery, S.K. Ozdemir,	J. Lee,	S. A. Maier, and M. S. Kim, \href{http://www.nature.com/nphys/journal/v9/n6/full/nphys2615.html}{Nature Physics \textbf{9}, 329–340 (2013)}.

\bibitem{Cuche2009}
A. Cuche, A. Drezet, Y. Sonnefraud, O. Faklaris, F. Treussart, J.-F. Roch, and S. Huant, \href{https://www.osapublishing.org/oe/fulltext.cfm?uri=oe-17-22-19969&id=187211}{Opt. Express \textbf{17}, 19969 (2009)}.

\bibitem{Cuche2010}
A. Cuche, O. Mollet, A. Drezet, and S. Huant, \href{http://pubs.acs.org/doi/abs/10.1021/nl102568m}{Nano Lett. \textbf{11}, 4566–4570 (2010)}.

\bibitem{Novotny2006} 
L. Novotny, and B. Hecht, \textit{Principles of Nano-Optics} (Cambridge University Press, Cambridge, England, 2006).

\bibitem{Chevalier2006} 
N. Chevalier, Y. Sonnefraud, J.F. Motte, S. Huant, and K. Karrai,  \href{http://scitation.aip.org/content/aip/journal/rsi/77/6/10.1063/1.2209950}{Rev. Sci. Instrum. \textbf{77}, 063704 (2006)}.

\bibitem{Berthel2015}
M. Berthel, Q. Jiang, C. Chartrand, J. Bellessa, S. Huant, C. Genet, and A. Drezet,\href{http://journals.aps.org/pre/abstract/10.1103/PhysRevE.92.033202}{Phys. Rev. E \textbf{92}, 033202 (2015)}.
\bibitem{Berthel2016}
M. Berthel, S. Huant, and A. Drezet \href{https://www.osapublishing.org/ol/abstract.cfm?uri=ol-41-1-37&origin=search}{Optics Letters \textbf{41}, 37-40 (2016)}.

\bibitem{Hecht1996}
B. Hecht, H. Bielefeldt, L. Novotny, Y. Inouye, and D. W. Pohl,\href{http://journals.aps.org/prl/abstract/10.1103/PhysRevLett.77.1889}{Phys. Rev. Lett. \textbf{77}, 1889 (1996)}.

\bibitem{Drezet2008Mater.Sci.Eng.B}
A. Drezet, A. Hohenau, D. Koller, A. Stepanov, H. Ditlbacher, B. Steinberger, F. R. Aussenegg, A. Leitner, and J. R. Krenn, \href{http://www.sciencedirect.com/science/article/pii/S0921510707005995}{Mater. Sci. Eng. B \textbf{149}, 220-229 (2008)}.

\bibitem{Drezet2013}
A. Drezet and C. Genet,\href{http://journals.aps.org/prl/abstract/10.1103/PhysRevLett.110.213901}{Phys. Rev. Lett. \textbf{110}, 213901 (2013)}.

\bibitem{Hohenau}
A. Hohenau, J. R. Krenn, A. Drezet, O. Mollet, S. Huant, C. Genet, B. Stein, and T. W. Ebbesen, \href{https://www.osapublishing.org/oe/abstract.cfm?uri=oe-19-25-25749}{Opt. Express \textbf{19}, 25749-25762 (2011)}.

\bibitem{Bliokh2008}
K. Y. Bliokh, Y. Gorodetski, V. Kleiner, and E. Hasman,\href{http://journals.aps.org/prl/abstract/10.1103/PhysRevLett.101.030404}{Phys. Rev. Lett. \textbf{101}, 030404 (2008)}.

\bibitem{Hosten2008}
O. Hosten, P. Kwiat, \href{http://science.sciencemag.org/content/319/5864/787}{Science \textbf{319}, 787−790 (2008)}.

\bibitem{Yin2013}
X. Yin, Z. Ye, J. Rho, Y. Wang, and X. Zhang, \href{http://science.sciencemag.org/content/339/6126/1405}{Science \textbf{339}, 1405−1407 (2013)}.

\bibitem{Dereux2003}
A. Dereux, C. Girard, C. Chicanne, G. Colas des Francs, T. David, E. Bourillot, Y. Lacroute, and J.C. Weeber, \href{http://iopscience.iop.org/article/10.1088/0957-4484/14/8/317/pdf}{Nanotechnology \textbf{14}, 935 (2003)}. 

\bibitem{Dung}
H. T. Dung, L. Kn\"{o}ll, and D. -G. Welsch, \href{http://journals.aps.org/pra/abstract/10.1103/PhysRevA.62.053804} {Phys.Rev. A \textbf{62}, 053804 (2000)}.

\bibitem{Baffou2008}
G. Baffou, C. Girard, E. Dujardin, G. Colas des Francs, and O. J. F. Martin,\href{http://journals.aps.org/prb/abstract/10.1103/PhysRevB.77.121101}{Phys. Rev. B \textbf{77}, 121101R (2008)}.

\bibitem{Krachmalnicoff}
V. Krachmalnicoff, D. Cao, A. Cazé, E. Castanié, R. Pierrat, N. Bardou, S. Collin, R. Carminati, and Y. De Wilde, \href{https://www.osapublishing.org/oe/abstract.cfm?uri=oe-21-9-11536}{Opt. Express \textbf{21}, 11536-11545 (2013)}.

\bibitem{submitted}
M. Berthel, Q. Jiang, A. Pham, J. Bellessa, C. Genet, S. Huant, A. Drezet, Submitted (2016).

\bibitem{Girard2005}
C. Girard, O. J. Martin, G. L\'{e}v\`{e}que, G. C. des Francs, and A. Dereux, \href{http://www.sciencedirect.com/science/article/pii/S000926140500076X}{Chem. Phys. Lett. \textbf{404}, 44 (2005)}.

\bibitem{Wang2014}
T. Wang, E. Boer-Duchemin, G. Comtet, E. Le Moal, G. Dujardin, A. Drezet, and S. Huant, \href{http://iopscience.iop.org/article/10.1088/0957-4484/25/12/125202?fromSearchPage=true}{Nanotechnology \textbf{25}, 125202 (2014)}.


\bibitem{Aouani2011}
H. Aouani, O. Mahboub, E. Devaux, H. Rigneault, T.W. Ebbesen, and J. Wenger\href{http://pubs.acs.org/doi/abs/10.1021/nl200772d}{Nano Lett. \textbf{11}, 2400 (2011)}.

\bibitem{Jun2011}
Y. C. Jun, K. C. Y. Huang, and M. L. Brongersma, \href{http://dx.doi.org/10.1038/ncomms1286}{Nature Com. \textbf{2}, 283 (2011)}.

\bibitem{Lodhal}
P. Lodhal, S. Mahmoodian, S. Stobbe, P. Schneeweiss, J. Volz, A. Rauschenbeutel, H. Pichler, P. Zoller, \href{https://arxiv.org/abs/1608.00446}{arXiv preprint arXiv:1608.00446}.




\end{thebibliography}
\end{document}